\documentstyle [epsfig, emulateapj]{article}

 1

\def\etal{{et al.~}}
\def\Hm{${\rm {H^-}}\,\,$}
\def\HH{${\rm {H_2}}\,\,$}
\def\HHp{${\rm {H_2^+}}\,\,$}
\def\erg{{\rm erg}}
\def\sr{{\rm sr}}

\def\cm{{\rm cm}}

\def\gs{\mathrel{\raise1.16pt\hbox{$>$}\kern-7.0pt
\lower3.06pt\hbox{{$\scriptstyle \sim$}}}}
\def\ls{\mathrel{\raise1.16pt\hbox{$<$}\kern-7.0pt
\lower3.06pt\hbox{{$\scriptstyle \sim$}}}}
\def\gtsima{$\; \buildrel > \over \sim \;$}
\def\ltsima{$\; \buildrel < \over \sim \;$}
\def\prosima{$\; \buildrel \propto \over \sim \;$}
\def\gsim{\lower.5ex\hbox{\gtsima}}
\def\lsim{\lower.5ex\hbox{\ltsima}}
\def\simgt{\lower.5ex\hbox{\gtsima}}
\def\simlt{\lower.5ex\hbox{\ltsima}}
\def\simpr{\lower.5ex\hbox{\prosima}}

\def\pp{\noindent\parshape 2 0truecm 17truecm 2truecm 15truecm}
\def\rf#1;#2;#3;#4 {\par\pp#1, #2, #3, #4. \par}

\def\pr{\ref@jnl{Phys.Rev}}     

\def\ie{{\frenchspacing\it i.e. }}

\def\href#1;#2 {{\bf #1} : {\em #2}}

\def\beq#1{\begin{equation}\label{#1}}
\def\eeq{\end{equation}}
\def\beqa#1{\begin{eqnarray}\label{#1}}
\def\eeqa{\end{eqnarray}}

\def\tento#1{\times 10^{#1}}

\def\s{{\rm \ s}}
\def\sr{{\rm \ sr}}

\def\erg{{\rm \ erg}}
\def\cm{{\rm \ cm}}

\def\kpc{{\rm \ kpc}}

\def\Hz{{\rm \ Hz}}

\def\HH{H$_2$ }
\def\H2p{H$_2^+$ }

\def\Hm{H$^-$ }

\def\mH2p{H_2^+}

\def\ka{{\kappa_\nu}}



\begin{document}
\thispagestyle{empty}
\title{  INTERGALACTIC \HH PHOTODISSOCIATION AND THE SOFT UV BACKGROUND\\
PRODUCED BY POPULATION III OBJECTS}  
 
\author {Benedetta Ciardi\altaffilmark{1}, Andrea Ferrara\altaffilmark{2,5}
and Tom Abel\altaffilmark{3,4}  }

\altaffiltext{1}{Universit\`a degli studi di Firenze, Dipartimento di
                    Astronomia, L.go E. Fermi 5, Firenze, Italy}
\altaffiltext{2}{Osservatorio Astrofisico di Arcetri, L.go E. Fermi 5, 
	  Firenze, Italy}
\altaffiltext{3}{ Laboratory for Computational Astrophysics, NCSA,
          University of Illinois at Urbana/Champaign, 405 N. Mathews
          Ave., Urbana, IL 61801} 
\altaffiltext{4}{Max-Planck-Institut f\"ur Astrophysik,
          Karl-Schwarzschild-Stra\ss e 1, 85748 Garching, Germany }
\altaffiltext{5}{Joint Institute for Laboratory Astrophysics, Campus Box 440,
          Boulder, CO}

\slugcomment{to be submitted to ApJ }
\received{---------------}
\accepted{---------------}

\begin{abstract}

	We study the effects of the ionizing and dissociating photons
	produced by Pop~III objects on the surrounding intergalactic medium.
	We find that the typical size of a \HH photodissociated region, 
        $R_d \approx
	1-5$ kpc, is smaller than the mean distance between 
	sources at $z \approx 20-30$, but
	larger than the ionized region by a factor depending on the 
	detailed properties of the emission spectrum. 
        This implies that clearing of intergalactic \HH occurs before
        reionization of the universe is complete. 
	In the same redshift range, the soft-UV background in the Lyman-Werner
        bands, when the intergalactic H and
	H$_{2}$ opacity is included, is found to be $J_{LW} \approx
	10^{-30}-10^{-27}$ erg cm$^{-2}$ s$^{-1}$ Hz$^{-1}$. This value 
	is well below the threshold required for the negative
	feedback of Pop~III objects on the subsequent galaxy formation
	to be effective in that redshift range. 

\end{abstract}

\section{Introduction}

At $z\approx 1100$ the intergalactic medium (IGM) is expected to
recombine and remain neutral until the first sources of ionizing radiation
form and reionize it.  \ion{H}{1} reionization is
essentially complete at $z\approx 5$, as it is evident by applying the 
Gunn--Peterson (1965) test to QSO absorption spectra. Recently,
observational evidences for a patchy \ion{He}{2} opacity of the IGM at
redshift $\approx 3$, have been collected. Although somewhat controversial 
(see Miralda-Escud\'e 1997 for a discussion), such inhomogeneous 
\ion{He}{2} absorption suggests an incomplete He reionization at these redshifts.  

Until recently, QSOs were thought to be the main source of
ionizing photons, but observational constraints suggest the existence of an early
population of pregalactic objects (Pop~III hereafter) which could
have contributed to the reheating, reionization and metal enrichment of
the IGM at high redshift. Indeed, the QSO population declines at
$z\simgt 3$ (Warren, Hewett \& Osmer 1994; Schmidt, Schneider \& Gunn 1995; 
Shaver \etal 1996), and it could not be sufficient to reionize the IGM at
$z\approx 5$ (Shapiro 1995 and references therein;
Madau 1998). In addition, metals have been now clearly detected 
in Ly$\alpha$ forest
clouds (Cowie \etal 1995; Tytler \etal 1995; Lu \etal 1998; Cowie \& Songaila
1998), 
suggesting that star formation
is ongoing at high redshift; finally, studies of the
[Si/C] abundance ratios in low column density absorption systems 
(Savaglio \etal 1997), also supported by theoretical work by Giroux \&
Shull (1997), show that they are consistent with  the presence of  
a softer, stellar component of the UV background.

In the following we will study the effects of the UV photons produced by
massive stars in Pop III objects on the surrounding IGM.
In order to virialize in the potential well of dark matter halos, 
the gas must have a mass greater
than the Jeans mass ($M_{b}>M_{J}$), which, at  $z\approx 30$, 
is $\approx 10^{5}
M_{\odot}$, corresponding to very low virial temperatures
($T_{vir}<10^{4}$ K). To have a further collapse and fragmentation of the gas,
and to ignite star formation, additional cooling is
required. It is well known that in these conditions the only efficient coolant
for a   plasma of primordial composition, is molecular hydrogen 
(Peebles \& Dicke 1968; Shapiro 1992 and references therein; 
Haiman \etal 1996; Abel \etal 
1997; Tegmark \etal 1997; Ferrara 1998). As the first 
stars form,
their photons in the energy range 11.26-13.6 eV are able to
penetrate the gas and photodissociate H$_{2}$ molecules both in the IGM
and in the nearest collapsing structures, if they can propagate that far
from their source. Thus, the existence of an UV
background below the Lyman limit due to Pop III objects (we will refer to it
 as ``soft-UV background'', SUVB), capable
of dissociating the H$_{2}$, could deeply influence subsequent 
small structure formation.
Haiman, Rees \& Loeb (1997, HRL), for example, have argued that Pop~III 
objects could depress the H$_{2}$ abundance in neighbor collapsing clouds, 
due to their UV photodissociating radiation, thus
inhibiting subsequent formation of small mass structures.

It is therefore important to assess the impact of these objects 
on their surroundings through detailed calculations of the various influence 
spheres, \ie ionization, photodissociation, and eventually also 
supernova metal enriched (Ciardi \& Ferrara 1997) spheres, produced
by Pop~IIIs. 
This approach will provide us with an insight of the basic aspects  
of the topology of ionization/dissociation in the early universe; moreover, in this way 
we hope to be able to ultimately understand
important issues as the reionization epoch and the initial stages of the 
galaxy formation in the universe. Obviously, the problem at hand      
depends  crucially on the mass, radiation spectrum and 
formation redshift of
pregalactic objects themselves. To clarify these dependences is the 
main aim of this paper.
In practice, we study the propagation of R-type ionization and 
dissociation fronts driven into the IGM by the radiation produced
by Pop~III massive stars in the early universe.  

In \S~2  we introduce the basic physical processes and give simple 
dimensional estimates; 
\S~3 defines the adopted model and the details of the radiation transfer
used in  the numerical work. 
The results and their implications are presented in \S 4  and a summary 
in \S~5 concludes
the paper.

\section{Physical processes  }

In this section we outline the basic physical processes underlying
the problem at hand. 
If massive stars form in Pop~III objects, their photons with $h\nu > 13.6$~eV
create a cosmological HII region in the surrounding IGM.
Its radius, $R_{i}$, can be estimated by solving the following standard equation
(Shapiro \& Giroux 1987) for the
evolution of the ionization front:
\begin{equation}
\frac{dR_{i}}{dt} -H R_{i}= \frac{1}{4 \pi n_{H} R_{i}^{2}} \; \left[
S_{i}(0) - \frac{4}{3} \pi R_{i}^{3} n_{H}^{2} \alpha^{(2)} \right];
\label{HIIrad}
\end{equation}
note that ionization equilibrium is implicitly assumed.
$H$ is the Hubble constant, $S_{i}(0)$ is the ionizing photon
rate, $n_{H}= 8\times 10^{-6} \Omega_b h^2 (1+z)^3$~cm$^{-3}$ is the
IGM hydrogen number density
and $\alpha^{(2)}$ is the hydrogen recombination rate to levels $\ge 2$.
First, we note that
since $R_i \ll c/H$ (see eq.~[\ref{rion}]), the cosmological expansion term $HR_i$ can be
safely neglected.
In its full form eq.~(\ref{HIIrad}) has been solved by Shapiro \& Giroux
(1987); such
solution is shown in Fig.~\ref{fig3} and discussed later on. 
If steady-state is assumed ($dR_i/dt \simeq 0$), then $R_i$ is approximately
equal to the Str\"omgren (proper) radius, $R_{S}=[3 S_{i}(0)/(4 \pi n_{H}^{2}
\alpha^{(2)})]^{1/3}$. In general, $R_{s}$ represents an upper limit 
for $R_{i}$, since the ionization front fills the time-varying
Str\"omgren radius only at very high redshift, $z \approx 100$ (Shapiro
\& Giroux 1987). For our reference parameters it is:
\begin{equation}
R_{i} \simlt R_{s} = 0.05 \, \left( {\Omega_b h^2} \right)^{-2/3}
(1+z)_{30}^{-2} \; S_{47}^{1/3}  \;\; \kpc,
\label{rion}
\end{equation}
where $S_{47}=S_{i}(0)/(10^{47}$ s$^{-1}$). The comparison between
$R_{i}$ and $R_{s}$ is also shown in Fig.~\ref{fig3}; $R_{s}$ is
typically 1.5 times larger than $R_{i}$ in this redshift range.

As mentioned above, in analogy with the cosmological HII region, photons
in the energy range 11.26 eV $\le h\nu \le$13.6 eV, create a 
photodissociated sphere in the surrounding IGM.
Lepp \& Shull (1984) have quantified the formation of primordial
molecules in the IGM after recombination.  In particular, primordial
\HH forms with a fractional abundance of $\approx 10^{-7}$ at redshifts
$\gsim 400$ via the \HHp formation channel. At redshifts $\lsim 110$
when the Cosmic Microwave Background radiation (CMB) is weak enough
to allow for significant formation of \Hm ions, even more \HH
molecules can be formed.  Due to the lack of molecular data, it has
unfortunately not been possible to follow the details of the \HHp
chemistry as its level distribution decouples from the CMB. Assuming
the rotational and vibrational states of \HHp to be in equilibrium
with the CMB, the \HHp photo--dissociation rate is much larger than
the one obtained by considering only photo--dissociations out of the
ground state.  Conservatively, one concludes that these two limits
constrain the \HH fraction to be in the range from
$10^{-6}-10^{-4}$. In fact both limits have been used in the
literature, see e.g. Lepp \& Shull (1984), Haiman \etal (1996),
Tegmark \etal (1997), Palla \etal (1995). In the following we will
assume that the \Hm channel for \HH formation is the dominant
mechanism, \ie that the \HHp photo--dissociation rate at high redshifts
is close to its equilibrium value. This leads to a typical value of
the primordial \HH fraction of $f_{H_{2}}\approx 2\tento{-6}$ (Shapiro
1992; Anninos
\& Norman 1996) which is found for model universes that satisfy the
standard primordial nucleosynthesis constraint $\Omega_b h^2 = 0.0125$
(Copi \etal 1995), where $\Omega_b$ is the baryon density parameter
and $H_0= 100 h$~km~s$^{-1}$~Mpc$^{-1}$ is the Hubble constant.

On average, $f_{d}\approx$~15\% of the \HH molecules that are radiatively excited by
photons in the Lyman-Werner (LW) bands decay to the continuum.  This so
called two-step Solomon process is the prime radiative \HH
dissociating mechanism in cosmological and interstellar enviroments.
Recently, Draine \& Bertoldi (1996) have studied this process in
great detail including line overlap and UV pumping and provide simple
fitting formulas to account for the self--shielding. 
Their treatment, however, cannot be directly applied to cosmological situations
mainly because of Hubble expansion that redshifts photons out of the LW lines. 
We treat \HH line radiation transfer in an expanding universe in detail 
in \S~4. 
The optically thin rate coefficient for the two-step photodissociation process 
is $k_{27} = 10^{8} J \s^{-1}$ (where $J$ is the average flux in the LW bands 
in units of $\erg\, \s^{-1} \cm^{-2} \Hz^{-1}$; the rate coefficients $k_{i}$ 
are labelled according to the nomenclature given in Abel \etal 1997). 

The main difference between ionization and dissociation spheres
evolution consists in the fact that there is no efficient mechanism to
re-form the destroyed H$_{2}$, analogous to H recombination. As a
consequence, it is impossible to define a photo-dissociation Str\"omgren
radius. However, given a point source that radiates $S_{LW}$ photons per 
second in the LW bands, an estimate of the maximum 
radius of the \HH photodissociated 
sphere, $R_{d}$, is the distance at which the (optically thin) 
photo--dissociation time
($\sim k_{27}^{-1}$) becomes longer than the Hubble time:
\begin{eqnarray}\label{rcrit}
R_{d} \simlt 2.5 \, h^{-1/2} (1+z)_{30}^{-3/4} S_{LW,47}^{1/2} \;\; \kpc,
\end{eqnarray}
where $S_{LW,47}=S_{LW}/(10^{47}$ s$^{-1}$).
Eqs.~(\ref{rcrit}) and~(\ref{rion}) show that the photodissociated
region
is larger than the ionized region; however, even if $S_{LW} \ll S_i(0)$
under most conditions $R_d$ cannot be smaller than $R_i$ since inside
the HII region, H$_2$ is destroyed by direct photoionization.
One might expect that
these sources are able to create a SUVB,
$J_{LW}$, even before all primordial \HH is dissociated. However, the
photo--dissociation time of molecular hydrogen will become shorter
than the Hubble time once this flux exceeds 
\begin{equation}
J^{crit}_{LW} = 6.2\tento{-4} J_{21} h  (1+z)_{30}^{3/2},
\end{equation}
where $J_{21}=10^{-21}\erg\s^{-1}\Hz^{-1}\cm^{-2}\sr^{-1}$. Values
of $J_{LW}>J^{crit}_{LW}$ will lead to
a substantial  destruction of primordial \HH and 
to a clearing of the universe in the LW bands. 

\section{Chemo-reactive fronts} \label{method}

To substantiate the above analytical estimates we have developed a
non--equilibrium multifrequency radiative transfer code to
study the detailed structure of R-type ionization and dissociation
fronts surrounding a
point source. We have adopted a standard CDM model ($\Omega_{m}$=1 and
$h=0.5, \sigma_8=0.6$), with a baryon density parameter $\Omega_{b}=0.06 \,
\Omega_{b,6}$,
of which a fraction $f_{b} \sim 0.08 \,f_{b,8}$ (Abel \etal 1997)
is able to cool and become available to form stars.
We study the evolution of ionization and dissociation fronts due to
photons with energy $h\nu >$ 13.6 eV and 11.26 eV $< h\nu <$ 13.6 eV
respectively, produced by a point source of baryonic mass $M_{b}=\Omega_{b} 
M \sim
10^{5} M_{\odot}$ ($M=10^{6} M_6~M_{\odot}$ is the total mass of the
object) forming at redshift $z=30$.
The program evolves the energy equation (see, for example, Shapiro, Giroux
\& Babul 1994; Ferrara \& Giallongo 1996) and the chemical network
equations (Abel \etal 1997), including 27 chemical
processes and 9 species (H, H$^-$, H$^+$, He, He$^+$, He$^{++}$, \HH,
\HHp and free electrons). The chemical abundances 
are initialized
according to the estimates provided by Anninos \& Norman (1996).
The cooling model includes collisional ionization, recombination,
collisional excitation and Bremsstrahlung cooling of atomic hydrogen and
helium, molecular hydrogen cooling, Compton cooling on the CMB, 
cosmological expansion cooling 
and all relevant heating mechanisms ({\it i.e.} photoionization and
photodissociation of all species).

\subsection{UV photon production}

We now estimate the UV photon production in
protogalactic objects. We assume that the total ionizing 
photon rate is proportional to 
the baryonic mass $M_{b}$: 
\begin{eqnarray}
S_{i}(0) & = & \frac{M_{b}}{m_{p} \; t_{OB} \; \tau} \; f_{uvpp} 
\; f_{esc} \; f_{b}  \nonumber \\
	 & \approx & 10^{47} f_{uvpp,48} f_{esc,20} \Omega_{b,6} f_{b,8}
M_{6}\;\;\; {\rm s^{-1}},
\label{photrate}
\end{eqnarray}
where $t_{OB}\approx 3\tento{7}$ yr is the average lifetime of massive stars;
$f_{uvpp} \approx 48000 \, f_{uvpp,48}$ is the number of 
UV photons produced per collapsed proton
(Tegmark \etal 1994); $f_{esc} \approx 0.2 \, f_{esc,20}$ is 
the photon escape fraction from the proto-galaxy; 
$\tau^{-1} \approx 0.6\% $ is the star
formation efficiency, normalized to the Milky Way. 
This simple estimate is within a factor of 2 of the value obtained from
the recently revised version of the Bruzual \& Charlot (1993, BC)
spectrophotometric code using
a Salpeter IMF, a burst of star formation, and a metallicity $Z=10^{-2}
Z_{\odot}$.
The adopted reference value $f_{esc}=0.2$ is an upper limit derived from
observational (Leitherer \etal 1995; Hurwitz \etal 1997) and
theoretical (Dove \& Shull 1994) studies.  

As a start, we have assumed a simple power-law spectrum 
above the Lyman limit, and a flux with constant intensity for photons
with energies lower than 13.6 eV:
\begin{equation}
j(\nu)= \left\{
\begin{array}{ll}
j_{0} \; \left( \frac{\displaystyle \nu}{\displaystyle \nu_{L}} \right)^{-\alpha}  &  13.6 \; {\rm eV} < h\nu < 850 \; {\rm
eV},\\
\beta \; j_{0} & h\nu < 13.6 \; {\rm eV},
\end{array}\right.
\label{jnu}
\end{equation}
where $\alpha=1.5$ (see \S~4 for a discussion on $\alpha$), and $j_{0}=S_{i}(0)
h/ \alpha \; {\rm erg \; s^{-1} \; Hz^{-1}}$ is the source luminosity
per unit frequency; $\beta=j(13.6^{-})/j(13.6^{+})$ is the ratio between the flux
just below and above the Lyman limit. The precise value of $\beta$ is rather
uncertain and can be estimated from the results of spectrophotometric
synthesis codes; in particular, we have used the BC code. We find that 
during most of the evolution of the stellar cluster the value of  $\beta$ 
remains close to 1; however, in the late evolutionary stages, 
after the massive stars producing the ionizing flux 
have died, $\beta$ increases up to $\approx 30$ (although in absolute value
the dissociating flux has decreased) since the dissociating photons are
partly produced by the remaining intermediate mass stars.
For this reason, the value of $\beta$ is somewhat dependent on the adopted IMF 
(throughout the paper we use a Salpeter IMF), which determines the relative
number of intermediate-to-massive stars.

\subsection{Radiative transfer}

The radiative transfer equation in an expanding universe is:
\begin{equation}
\frac{1}{c} \frac{\partial{J_{\nu}}}{\partial{t}} + \frac{1}{a}
\frac{\partial{J_{\nu}}}{\partial{R}} -\frac{H}{c} \left(
\frac{\partial{J_{\nu}}}{\partial{\nu}} \nu - 3 J_{\nu} \right) =
\epsilon_{\nu} - \ka J_{\nu},
\label{trans}
\end{equation}
where $J_{\nu}(r)=j(\nu)/R^{2}$ [erg s$^{-1}$ cm$^{-2}$ Hz$^{-1}$]; 
$a=(1+z_{em})/(1+z_{abs}) \sim 1$, with $z_{em}$ and $z_{abs}$
emission and absorption redshift respectively.
As the scales of interest are small, eq.~(\ref{trans}) can be used in the
local approximation, \ie we can neglect the cosmological redshift and the time
dependent terms; 
$\epsilon_{\nu}$ [${\rm erg \; s^{-1} \; cm^{-3} \;Hz^{-1}}$],
the emissivity of the gas, can be
neglected when using the canonical ``on the spot approximation'' (cf.
Osterbrock 1989); $\ka[\cm^{-1}]$, the frequency dependent absorption
coefficient, is implicitly assumed to vary on time scales larger 
then the photon crossing time. Under these approximations, 
eq.~(\ref{trans}) reduces to:
\begin{equation}
\frac{\partial{J_{\nu}}}{\partial{R}} = - \ka J_{\nu}.
\label{trans1}
\end{equation}
In the expression for $\ka= \sum_{i} \sigma_{i} n_{i}$, with $\sigma_{i}$
[cm$^{2}$] and $n_{i}$ [cm$^{-3}$] respectively the cross section and number
density of species $i$, we have included the contribution of all the
relevant species.

\section{Results and Implications}

Using the above assumptions, we have derived the redshift evolution of the IGM temperature and chemical
abundances as a function of distance $R$ from the central source, which is
supposed to turn on at $z=30$. 

\begin{figure}[t]
\centerline{\psfig{figure=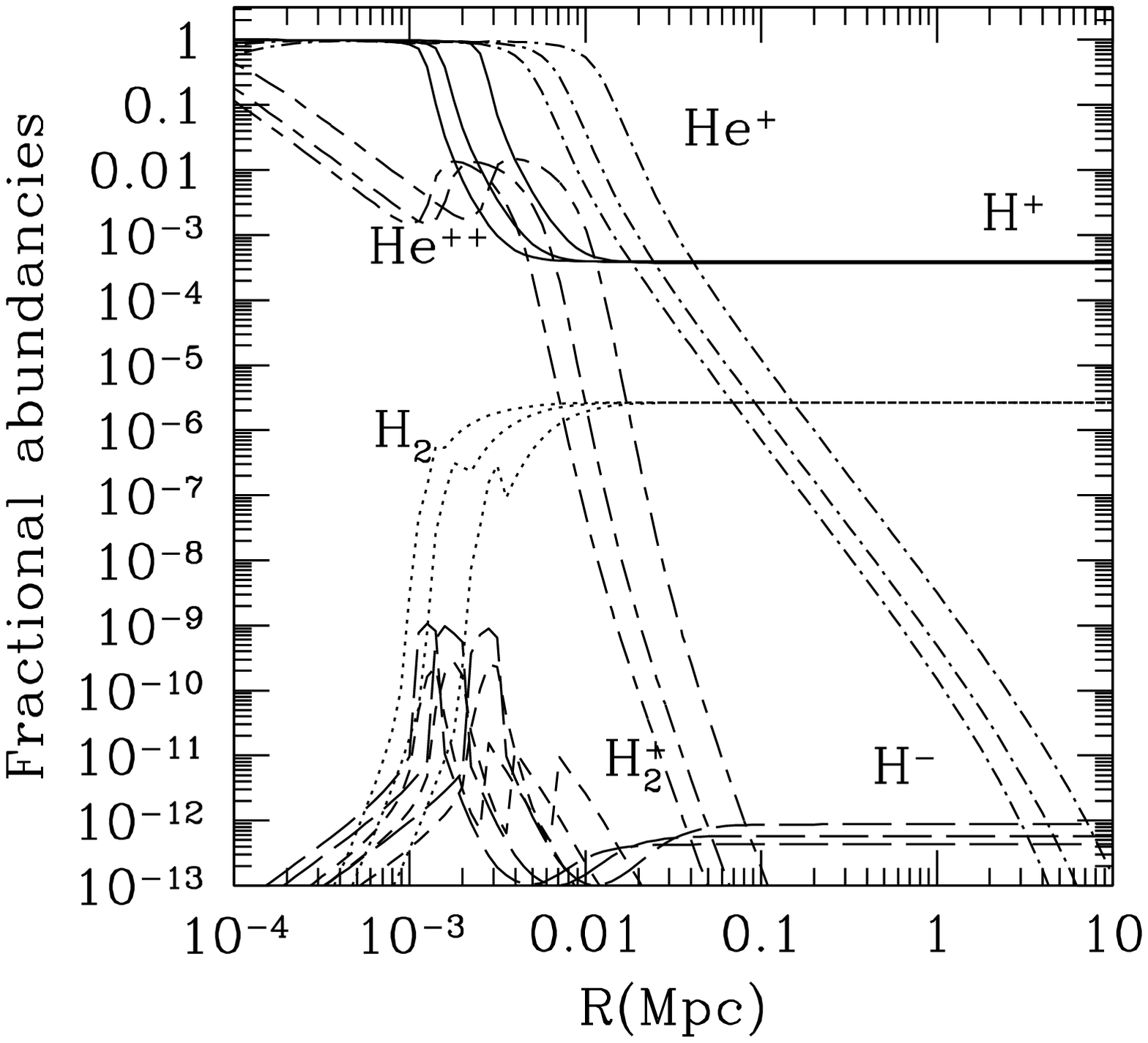}}
\caption{\label{fig1}{Evolution of chemical abundancies as a function of distance
from a Pop~III of total mass $M \approx 10^{6} \, M_{\odot}$, turning on
at $z \approx 30$. From left to right the curves refer to $z = 27, 23,
19$.}}
\end{figure}

\begin{figure}[t]
\centerline{\psfig{figure=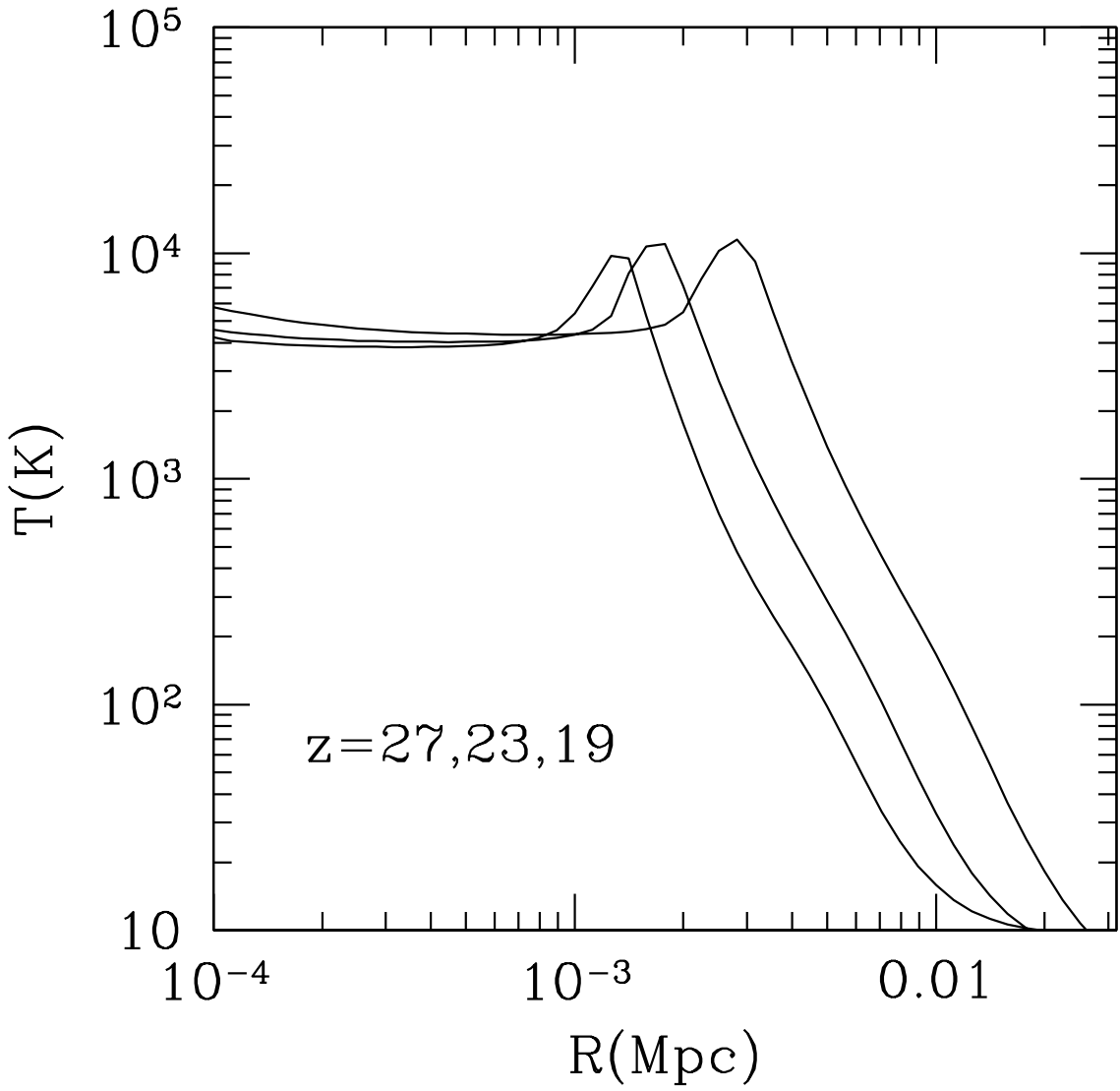}}
\caption{\label{fig2}{Evolution of gas temperature as a function of distance
from a Pop~III of total mass $M \approx 10^{6} \, M_{\odot}$, turning on
at $z \approx 30$. From left to right the curves refer to $z = 27, 23,
19$.}}
\end{figure}

\begin{figure}[t]
\centerline{\psfig{figure=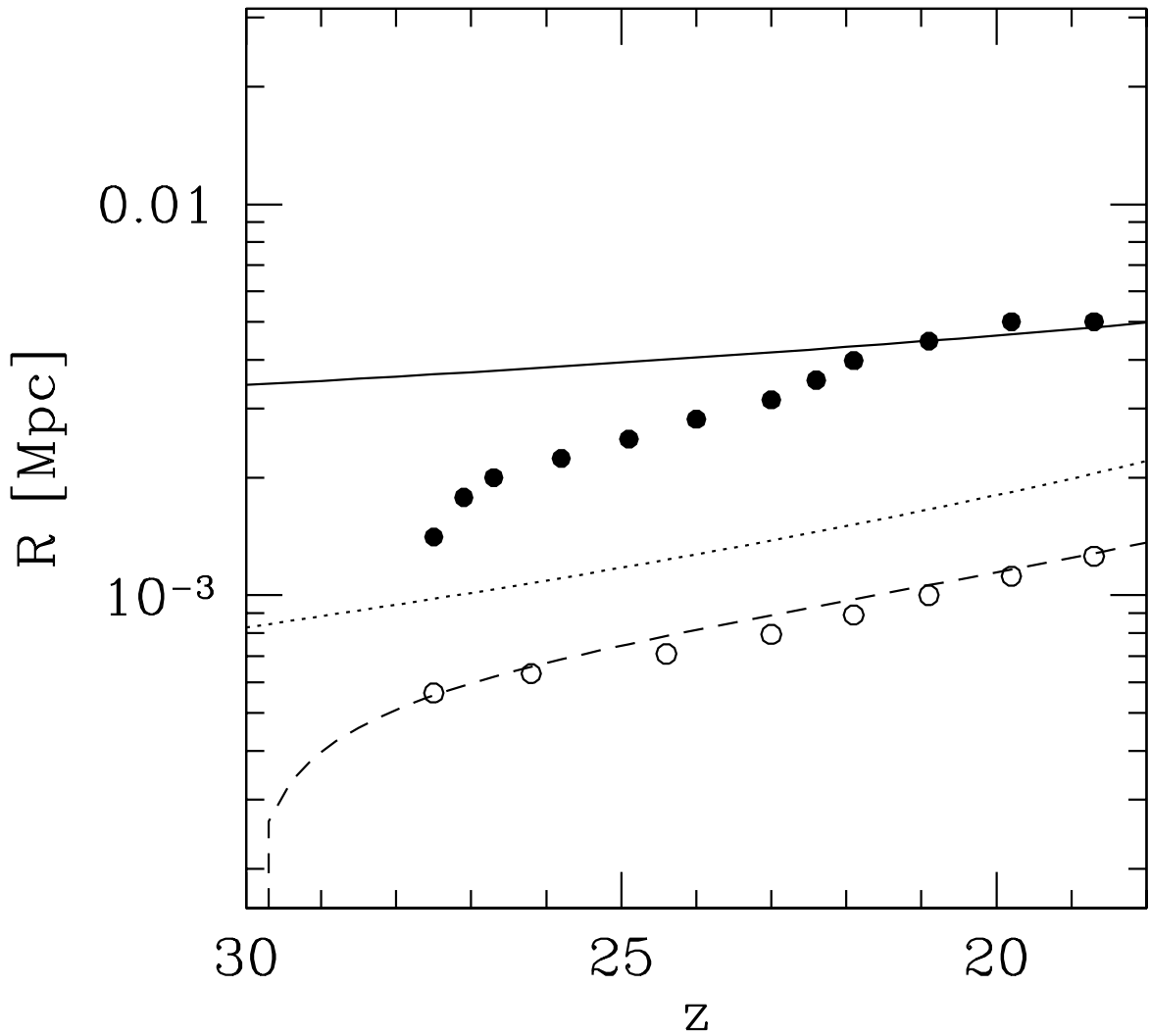}}
\caption{\label{fig3}{Ionization radius, $R_i$ (open circles) and 
photodissociation radius, $R_d$ (filled)
of the regions produced by a Pop~III of total mass $M \approx 10^{6} \,
M_{\odot}$, turning on at $z \approx 30$, as a function of redshift.
Also shown is the maximum radius of the dissociated region (solid line),
given by eq.~(\ref{rcrit}), the Str\"omgren radius $R_{s}$ (dotted), given by
eq.~(\ref{rion}), and the solution of eq.~(\ref{HIIrad})
(dashed).}}
\end{figure}

In Fig.~\ref{fig1} we show these results for the specific value of the 
parameter 
$\beta=1$ (see eq.~[\ref{jnu}]), which should appropriately describe the
radiation spectrum in the early stages of the evolution of the Pop~III
stellar cluster, as discussed in \S 3; Fig.~\ref{fig2} illustrates the analogous 
temperature evolution (which is independent on the value of $\beta$, since
photodissociation is negligible with respect to photoionization 
in terms of heating). For a comparison with previous works, see for
example Shapiro, Giroux \& Kang (1987).

The location  of the ionization and dissociation fronts, moving away from 
the central radiation source, is clearly identified by the sudden drop
in the ionized hydrogen and raise of the molecular hydrogen 
abundance, respectively. 
As the dissociating flux is diluted at large $R$, the H$_{2}$ abundance 
reaches its asymptotic value, given by the IGM relic fraction at recombination
$\approx 2\times 10^{-6}$.
In addition, the He singly ionized      
region is also shown to be larger than the H one, a well known effect produced
by the penetrating high energy photons of the power-law radiation spectrum. 
A similar feature is not seen in He$^+$ ionization due to the paucity of
photons capable to ionize such a species. 
Inside the dissociated sphere, \HH is basically completely destroyed
by the two-step Solomon process and by ionization.
H$^{-}$ and H$_{2}^{+}$ ions abundancies have a peak at the same location as
the temperature bump (cf. Fig.~\ref{fig2}); this temperature increase, together
with the persisting large supply of free electrons, favors the
formation rate of the above species;
nevertheless, the \HH formation rate is completely negligible. 
The temperature (Fig.~\ref{fig2}) inside the ionized
region is roughly constant and equal to $\approx 10^{4}$ K; the small bump
in the temperature profile is due to the decrease in the number of
electrons responsible for collisional excitation of atomic cooling lines,
in turn due to the decrease of the ionizing flux.
The shape of the curves depends on
the choice of the flux power law index, $\alpha$: in addition to the
standard case $\alpha=1.5$, we have also run the case of softer spectra,
$\alpha = 5$. In this case 
the H ionization transition region becomes steeper and
the He ionized regions are shrinked. These differences are, however,  not  
particularly relevant to the main focus of this paper, since the H$_2$
abundance is not sensibly affected.

\begin{figure}[t]
\centerline{\psfig{figure=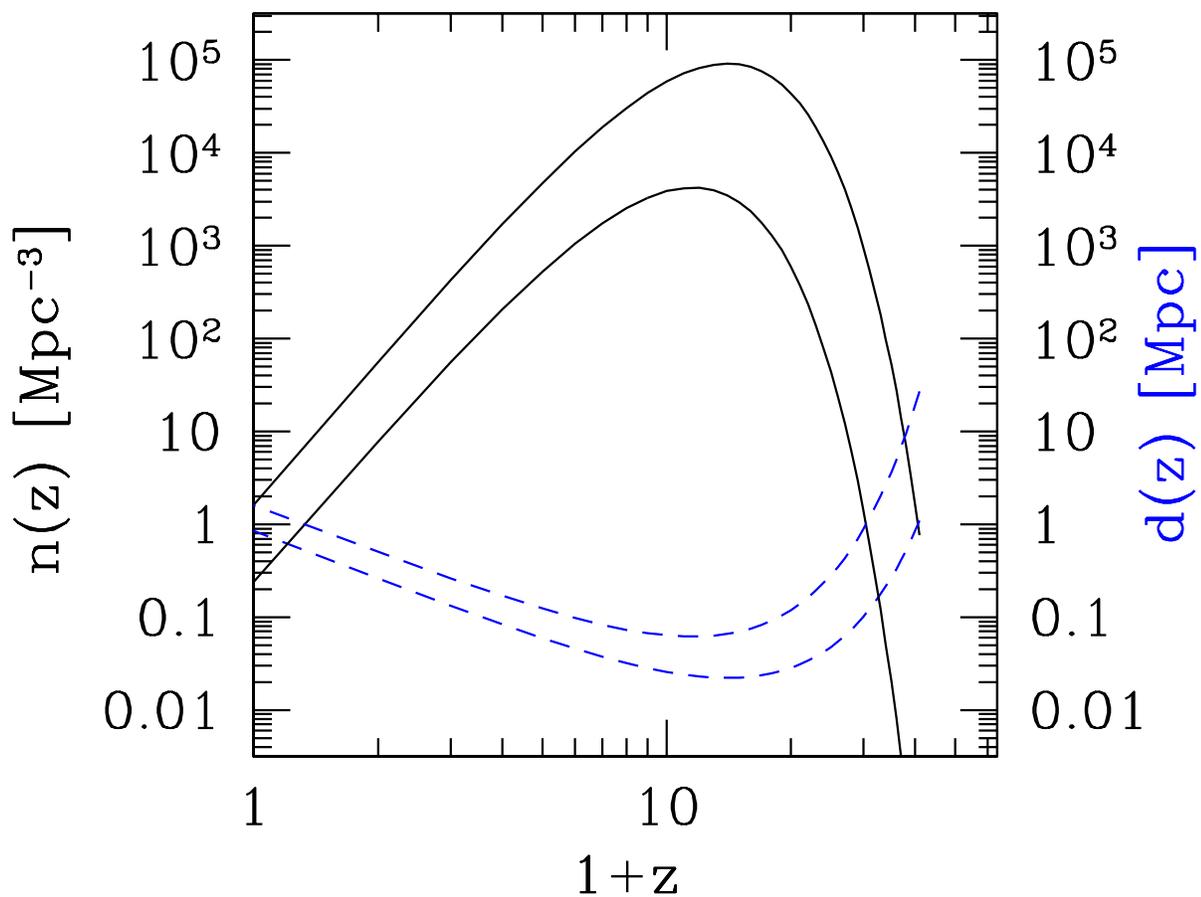}}
\caption{\label{fig4}{Evolution of the proper number density $n(z)$ (solid lines) and
typical interdistance $d(z)$ (dashed) of dark matter halos with circular
velocity $v_{c} \approx 7$ km s$^{-1}$ (upper solid line and lower
dashed) and $\approx$ 15 km s$^{-1}$ (lower solid line and upper dashed), 
corresponding to mass $M \approx 10^{6} M_{\odot}$ and $\approx 10^{7} M_{\odot}$
respectively,
in the redshift range of interest 20-30, in a $\sigma_{8}=0.6$ CDM model.}}
\end{figure}

In addition to the case $\beta=1$ we have also run cases with $\beta$ varying
in the range 1-100, finding that the above
general considerations remain essentially unchanged. 
The only important change is represented by the absolute and relative values
of the dissociation, $R_d$, and ionization, $R_i$, radii, that we are discussing
next.
Practically, we have defined $R_{i}$ and
$R_{d}$ as the radius   at which the ionized fraction deviates from unity
and the H$_2$ fraction differs from its asymptotic value by
less than an arbitrary constant $p\approx$ few percent, respectively.
This choice leads to a good agreement with the analytical solution
of eq.~(\ref{HIIrad}) for $R_i$.
In Fig.~\ref{fig3} we plot the numerical values of $R_{i}$ and 
$R_{d}$ as function of redshift, for
$\beta=1$. Eq.~(\ref{rcrit}) with $S_{LW}=\beta
S_{i}(0)$ gives a simple estimate for the asymptotic value of $R_{d}$.
This limiting value is reached after about one photo-dissociation time
$t_{d}$. For $\beta=1$ it is  $R_d \simeq 3.5 R_i$, 
while for higher values of $\beta$, 
$R_{d}$ becomes much larger than $R_{i}$, because of the increased
number of LW photons.
This roughly agrees with the estimate obtained by taking the ratio of the two
radii from eqs.~(\ref{rcrit}) and~(\ref{rion}):
\begin{eqnarray}
\label{ratio}
{R_{d}\over R_i} = 6 \, h^{5/6} \Omega_{b,6}^{2/3} (1+z)_{30}^{5/4}\beta^{1/2} 
S_{47}^{1/6}.
\end{eqnarray}
Since H$_{2}$ inside the ionized region is always destroyed via direct
photoionization, by
definition it must be $R_{d} \ge R_{i}$; then eq.~(\ref{ratio}) holds
down to a redshift $1+z \approx 7 h^{-2/3} \Omega_{b,6}^{-8/15}
\beta^{-2/5} S_{47}^{-2/15}$.  As we will see, it is important to
compare the size of the dissociated regions around Pop III objects
with their average interdistance, $d$, to determine if the surviving
intergalactic \HH can build up a non-negligible optical depth to LW
photons.  The proper number density distribution, $n(v_c,z)$, of dark matter
halos as function of their circular velocity, $v_c$, and redshift can be
computed by using the Press-Schechter formalism (Press \& Schechter
1974; for a modern formulation see, for example, White \& Frenk 1991
and Bond \etal 1991). For the adopted CDM model 
such distribution is shown in Fig. \ref{fig4} for
the values $v_c = 7$~km~s$^{-1}$ (15 km s$^{-1}$), that correspond to
a mass $M \approx 10^6 M_\odot$ ($10^7 M_\odot$) in the redshift
interval 20-30. Also shown is the evolution of the
typical interdistance between such objects, $d(z)\simeq
[n(v_c,z)]^{-1/3}$.  We find that the mean distance between Pop III
objects in the relevant redshift range is equal to 0.01-1 Mpc.  As $d$
is bigger than the typical derived size of H$_{2}$ regions at these
redshifts (see eq.~[\ref{rcrit}]), the H$_{2}$ regions can hardly 
overlap and completely destroy the primordial \HH molecules.
From these simple considerations, one can conclude that \HH 
photodissociated sphere overlapping will become important at $z 
\simlt  20$. This finding is also supported by numerical simulations
that will be presented in a forthcoming communication.
Note that eq. \ref{ratio} also implies that clearing of intergalactic
\HH occurs before reionization of the universe is complete.

\subsection{Soft-UV background}

In the redshift range 20-30, before overlapping becomes substantial, 
a SUVB will be present.
To derive the intensity of the SUVB, $J_{LW}$, we assume that
the filling factor of the dissociated regions is small, as it
follows from the inequality $R_{d} \ll d$. 
In order to evaluate the SUVB properly including the intergalactic 
H$_{2}$ attenuation, the effects of cosmological expansion must be 
taken into account. To this aim it is 
necessary to make a detailed treatment of the radiative transfer through 
LW H$_{2}$ lines. 
LW lines are optically thin in the redshift range of interest, 
with $\tau_{i}=N_{H_{2},i}\sigma_{i}\sim 0.05$, essentially for all the lines. This
implies that as $J_{LW}$ is redshifted due to the cosmological expansion, it is
attenuated by each LW line encountered by a factor $e^{-\tau_{i}}$.
Also, for an IGM temperature of about 10 K (adiabatic expansion 
can make the neutral, post recombination IGM cooler than the CMB), 
the Doppler line width is $\sim 1.25 \tento{-5}$ eV, thus implying that 
line overlapping can be safely neglected.
Abgrall \& Roueff (1989) have included in their study of
classic H$_2$ PDRs more than 1000 LW lines. As in our case H$_2$
formation which leaves the molecule in excited roto/vibrational levels, 
is negligible, a smaller number of lines $\approx 70$ -- involving the ground state only -- needs to be considered.
Globally, $J_{LW}$ is
attenuated by a factor $e^{-\tau_{H2}}$,
where $\tau_{H2}=\sum_{i} \tau_{i}$, and $i$ runs up to the 71 lines
considered from the ground roto/vibrational states; we obtain a $\tau_
{H_{2}} \simlt 3$, depending on the number
of lines encountered, and thus on the photon energy. 
In general, one can thus derive the intensity of the background as:
\begin{equation}
J_{LW}(\nu,z)=c \int^{z}_{z_{on}} \epsilon(\nu',z') \; {\rm
e}^{-\tau_{H2}(\nu',z')}
\frac{(1+z)^{3}}{(1+z')^{3}} \left\vert \frac{dt}{dz'}\right\vert dz',
\label{back}
\end{equation}
\begin{equation}
\epsilon(\nu',z')=\int^{v_{max}}_{v_{min}} j(\nu') {\cal N}(v_{c},z') dv_{c},
\end{equation}
where $z_{on}=30$, $v_{min}=7$ km s$^{-1}$, $v_{max}$=15 km s$^{-1}$,
$\nu'=\nu (1+z')/(1+z)$; 
$\epsilon(\nu',z')$ [erg cm$^{-3}$ s$^{-1}$ Hz$^{-1}$] is the proper emissivity due
to sources with $v_{min} \le v_{c} \le v_{max}$, corresponding to
objects in which H$_{2}$ cooling is dominant; $j(\nu')$ is given in
eq.~(\ref{jnu}) and ${\cal N}(v_{c},z')$ is the source number density in
the velocity interval $dv_{c}$. The value of $J_{LW}$
is not sensitive to different choices of $v_{max}$, as there are very
few bound structures with $v_c > v_{max}$ in the redshift range of
interest; on the other hand, lower values of $v_{min}$ have little
effect on $J_{LW}$, due to the faintness of these low mass objects.

\begin{figure}[t]
\centerline{\psfig{figure=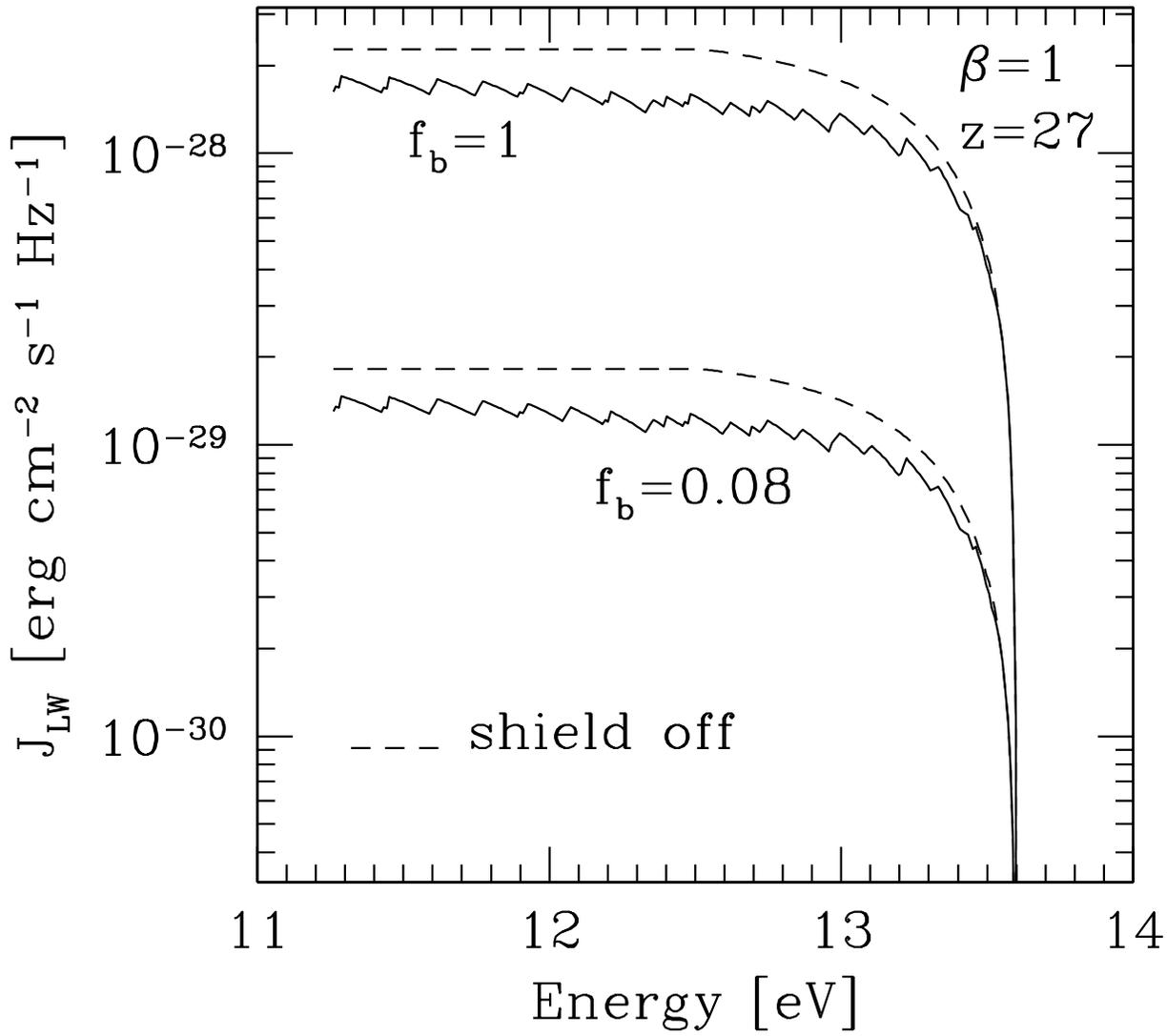}}
\caption{\label{fig5}{Spectrum of the SUVB,
$J_{LW}$, for $z=27$, $\beta=1$ and for baryon cooling efficiency
$f_{b}=0.08, 1$; the intergalactic H$_{2}$ Lyman-Werner
opacity is either included (solid lines) or neglected (dashed).}}
\end{figure}

In Fig.~\ref{fig5} the spectrum of the SUVB at $z=27$
is shown, for different choices of the baryon cooling efficiency $f_{b}=0.08,1$
(which in turn implies different $j_{0}$ values) and in which the IGM 
LW attenuation is either included or neglected for comparison sake. 
As $J_{LW}$ depends linearly on  $\beta$, the results for different values
of $\beta$ can be easily obtained by appropriately scaling the plotted curves. 
From Fig.~\ref{fig5} we see that the intergalactic H$_{2}$ absorption 
decreases the SUVB intensity by approximatly 30\%.

\begin{figure}[t]
\centerline{\psfig{figure=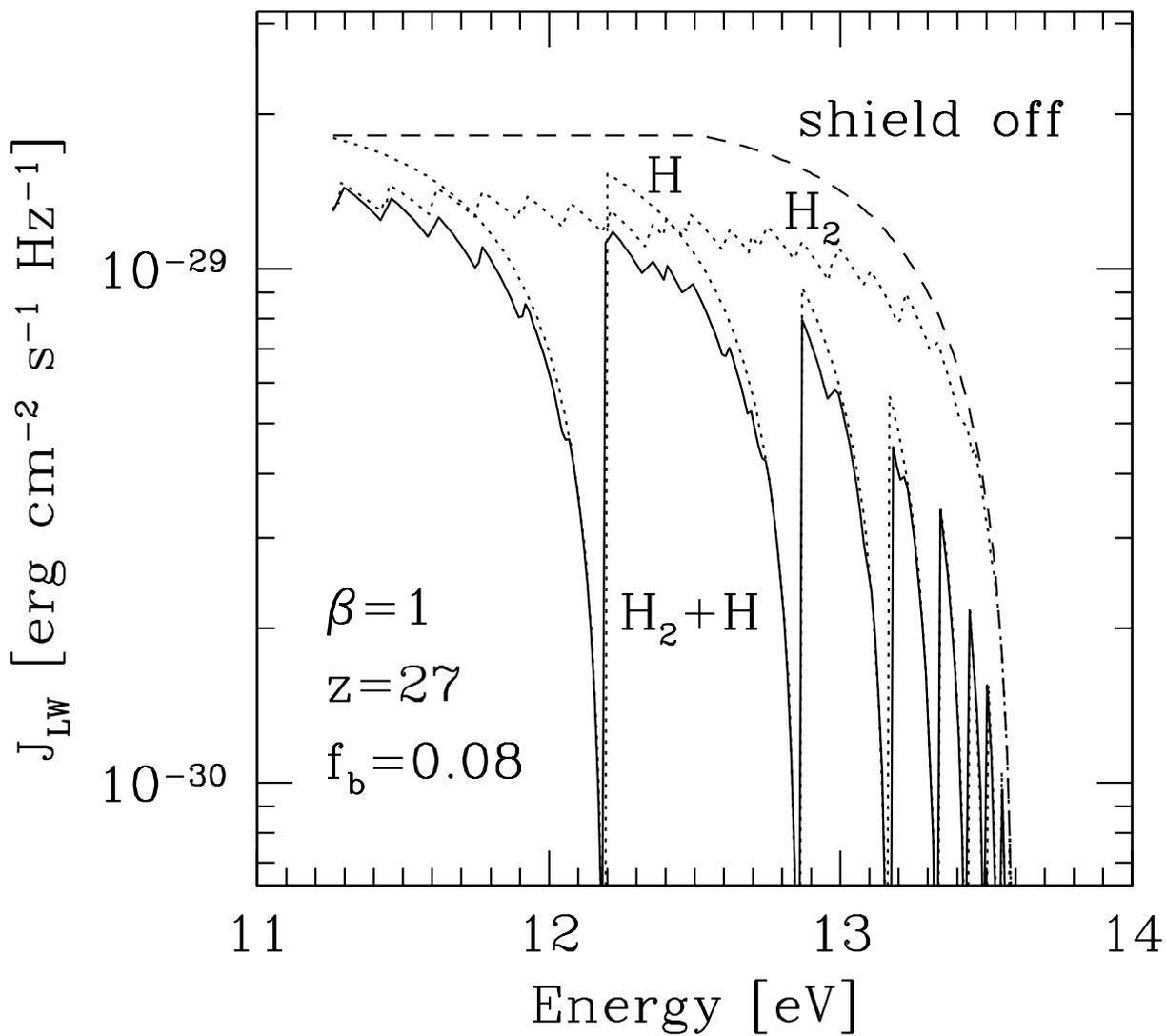}}
\caption{\label{fig6}{Same as Fig. 5 for the particular values
$z=27$, $\beta=1$ and baryon cooling efficiency
$f_{b}=0.08$; a comparison is shown between four different prescriptions
for the intergalactic attenuation: no shielding 
(dashed line), neutral H lines opacity only (dotted), H$_{2}$ lines opacity 
only (dotted) and the sum of 
H$_{2}$ and neutral H lines opacity (solid).}}
\end{figure}

In addition to the LW absorption lines, we have also investigated the 
relative importance of the neutral H lines absorption in the
calculation of $J_{LW}$, in analogy with the study of HRL. 
In Fig.~\ref{fig6} we compare the effect of such lines and H$_{2}$ lines
on the SUVB attenuation, for a
typical choice of the parameters. The H lines are optically thick at
their center; this, combined with the effect of the cosmological
expansion, produces the typical sawtooth modulation of the spectrum.
On the other hand, the radiative decay of the excited H atoms produces
Ly$\alpha$ and other Balmer or lower line photons, which are out of the LW
energy range and result in an increase of the flux just
below the Ly$\alpha$ frequency. From Fig.~\ref{fig6} we see that the
H line attenuation dominates on the H$_2$ one over the all range of
frequencies, except from the limited spectral regions blueward of the 
Ly$\beta$ and Ly$\gamma$ resonances.  

\begin{figure}[t]
\centerline{\psfig{figure=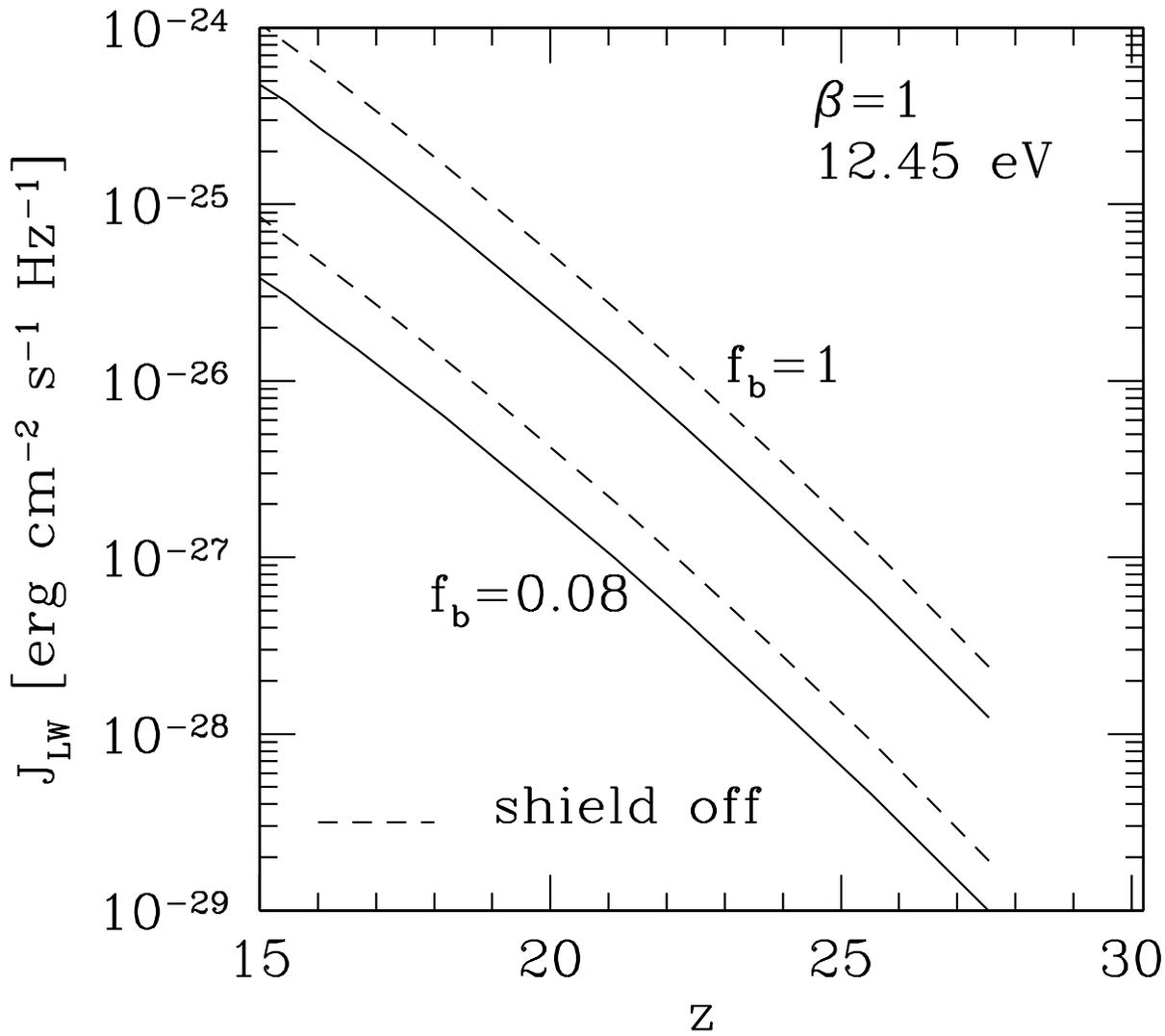}}
\caption{\label{fig7}{Evolution with redshift of the SUVB, 
$J_{LW}$, for $h\nu=12.45$ eV, $\beta=1$ and for baryon cooling efficiency
$f_{b}=0.08, 1$; the intergalactic H lines and H$_{2}$ Lyman-Werner
opacity is either included (solid lines) or neglected (dashed).}}
\end{figure}

The evolution of $J_{WL}$ with $z$ is shown in Fig.~\ref{fig7}, for 
$\langle h\nu\rangle$=12.45 eV, the central frequency of the LW band. 
From the plot we see that tipically, a SUVB intensity
$J_{LW} \approx 10^{-30}-10^{-26}$ erg cm$^{-2}$ s$^{-1}$ Hz$^{-1}$
is produced by Pop~III objects.

These results are particularly important when the effects of the
possible "negative feedback" are to be considered. HRL concluded that
in principle a SUVB created by pregalactic objects, would be able to
penetrate large clouds, and, by suppressing their \HH abundance,
prevent the collapse of the gas. One of their conclusions (HRL Erratum
1997) is that in the
redshift range $z=25-35$, a minimum SUVB intensity, $J_{LW} \approx 
10^{-24}$ erg cm$^{-2}$ s$^{-1}$ Hz$^{-1}$ is required to
significantly affect or inhibit the collapse of forming objects.
From Fig. 7 we conclude that at these redshifts the intensity of the SUVB is
well below the threshold required for the negative
feedback of Pop~III objects on the subsequent galaxy formation to be
effective. 
Clearly, if at redshift $\approx 20$ complete overlapping of photodissociated 
regions occurs, as previously suggested, the SUVB intensity can be increased to
interesting values for negative feedback effects. However, by that time
a considerable fraction of the objects in the universe that must rely on 
\HH cooling (\ie $v_c \simlt 15$~km~s$^{-1}$) for collapse might be already in 
an advanced evolutionary stage (see Fig. 4) and actively 
forming stars. It follows that negative feedback can at best only
partially influence the formation of small objects.
To confirm this hypotyhesis, that depends on the details of structure 
formation, numerical simualtions are required.  

\section{Summary}\label{soft}

We have studied the evolution of ionization and
dissociation spheres produced by and surrounding Pop~III objects,
which are supposed to have total masses $M \approx 10^{6} M_{\odot}$ 
and to turn on their radiation field at a redshift $z=30$.
By a detailed numerical modelling of non-equilibrium radiative transfer,
we find that the typical size of the dissociated region is $R_d \approx 1-5
$ kpc, while
the ionization radius is about 3.5 times smaller; this result is in good
agreement with the analytical estimate of eq.~(\ref{rcrit}).  
The mean distance
between Pop~IIIs at $z \approx 20-30$ in a CDM model, $d \approx 0.01-1$ Mpc, 
is larger than $R_{d}$. 
Thus, at high redshift, the
photodissociated regions are not large enough to overlap. In the same
redshift range, the soft-UV
background in the LW bands when the intergalactic H and
H$_{2}$ opacity is included, is found to be $J_{LW} \approx
10^{-30}-10^{-27}$ erg cm$^{-2}$ s$^{-1}$ Hz$^{-1}$, depending on the
star formation efficiency; this value 
is well below the threshold required for the negative
feedback of Pop~III objects on the subsequent galaxy formation
to be effective before redshift $\approx 20$.

\acknowledgments T.A. acknowledges support from NASA grant NAG5-3923.
We thank F. Bertoldi, E. Corbelli and P. Madau for useful discussions;
M. Rees, N. Gnedin and the referee P. Shapiro for their stimulating comments.

\newpage

\vfill
\eject
\end{document}